\documentclass[traditabstract]{aa}
\usepackage{graphicx}

\def\gtsima{$\; \buildrel > \over \sim \;$}
\def\ltsima{$\; \buildrel < \over \sim \;$}
\def\gtrsim{\lower.5ex\hbox{\gtsima}}
\def\lesssim{\lower.5ex\hbox{\ltsima}}

\begin{document}

\title{A cosmological view of extreme mass-ratio inspirals in nuclear star clusters}

\author{M. Mapelli\inst{1}, E. Ripamonti\inst{2}, A. Vecchio\inst{3}, Alister W. Graham\inst{4}, A. Gualandris\inst{5}}
\institute{INAF-Osservatorio Astronomico di Padova, Vicolo dell'Osservatorio 5, I--35122, Padova, Italy\\ \email{michela.mapelli@oapd.inaf.it}
\and
Universit\`a Milano Bicocca, Dipartimento di Fisica G.Occhialini, Piazza delle Scienze 3, I--20126, Milano, Italy
\and
School of Physics and Astronomy, University of Birmingham, Edgbaston, Birmingham B15 2TT, UK
\and
Centre for Astrophysics and Supercomputing, Swinburne University of Technology, Hawthorn, Victoria 3122, Australia
\and
Max-Planck Institut f\"ur Astrophysik, Karl-Schwarzschild-Strasse 1, D-85748 Garching, Germany
}
\titlerunning{A cosmological view of EMRIs in NCs}
\authorrunning{Mapelli et al.}
 
\abstract{There is increasing evidence that many galaxies host both a nuclear star cluster (NC) and a super-massive black hole (SMBH). Their coexistence is particularly prevalent in spheroids with stellar mass $10^{8}-10^{10}$ M$_\odot{}$. 
We study the possibility that a stellar-mass black hole (BH) hosted by a NC inspirals and merges with the central SMBH.
Due to the high stellar density in NCs, extreme mass-ratio inspirals (EMRIs) of BHs onto SMBHs in NCs may be important sources of gravitational waves (GWs).
 We consider sensitivity curves for three different space-based GW laser interferometric mission concepts: the Laser Interferometer Space Antenna (LISA), the New Gravitational wave Observatory (NGO) and the DECi-hertz Interferometer Gravitational wave Observatory (DECIGO). We predict that, under the most optimistic assumptions, LISA and DECIGO will  detect up to thousands of EMRIs in NCs per year,
 while NGO will  observe up to tens of EMRIs per year. 
We explore how a number of factors may affect the predicted rates.
In particular, if we assume that the mass of the SMBH scales with the square of the host spheroid mass in galaxies with NCs, rather than a linear scaling, then the event rates are more than a factor of 10 lower for both LISA and NGO, while they are almost unaffected in the case of DECIGO.}{}{}{}{}
\maketitle

%

\section{Introduction}

Nuclear star clusters (NCs) are massive star clusters located at the centre of galaxies (see, e.g., B\"oker 2010 for a recent review). They are common and quite ubiquitous, as the fraction of galaxies with unambiguous NC detection is $\sim{}75$ per cent in late-type (Scd-Sm) spirals (B\"oker et al. 2002), $\sim{}50$ per cent in earlier-type (Sa-Sc) spirals (Carollo et al. 1997) and $\sim{}70$ per cent in low and intermediate luminosity early-type galaxies (Graham \&{} Guzm\'an 2003; C$\hat{\rm o}$t\'e et al. 2006). The typical size of NCs is similar to that of Galactic globular clusters (GCs, half-light radius $\approx{}2-5$ pc, Geha, Guhathakurta \&{} van der Marel 2002; B\"oker et al. 2004; C$\hat{\rm o}$t\'e et al. 2006), but NCs are $1-2$ orders of magnitude brighter (and more massive) than Galactic GCs (Walcher et al. 2005). Unlike most of the Galactic GCs, NCs have often a complex star formation (SF) history, with multiple episodes of SF (Rossa et al. 2006; Walcher et al. 2006). NCs are of course not the Bahcall-Wolf (Bahcall \&{} Wolf 1976, 1977) cusps predicted to exist around SMBHs: NC sizes are much larger (e.g. Merritt \&{} Szell 2006) and they are observed in galaxies thought not to host SMBHs (e.g. Ferrarese et al. 2006). 

NCs obey scaling relations with host-galaxy properties (such as galaxy mass, spheroid luminosity and velocity dispersion, e.g., Balcells et al. 2003;  Graham \&{} Guzm\'an 2003; Balcells, Graham \&{} Peletier 2007; Ferrarese et al. 2006; Wehner \&{} Harris 2006; Rossa et al. 2006; Graham \&{} Driver 2007). These scaling relations are similar to those observed for super-massive black holes (SMBHs), suggesting that there is a link between NCs and SMBHs. Various studies (Ferrarese et al. 2006; Wehner \&{} Harris 2006) indicate that SMBHs (NCs) are found predominantly above (below) a stellar mass threshold $M_\ast{}\approx{}10^{10}\,{}{\rm M}_\odot$. However, the presence of a SMBH and that of a NC do not seem to be mutually exclusive. In addition to the Milky Way (MW, see, e.g., Sch\"odel et al. 2007, 2009), Filippenko \&{} Ho (2003) identified at least one galaxy (NGC~4395) hosting both a SMBH and a NC, and Graham \&{} Driver (2007) subsequently reported the existence of two additional such galaxies (NGC~3384 and NGC~7457). The sample of galaxies hosting both a SMBH and a NC was substantially increased by  Gonzalez Delgado et al. (2008, 2009), Seth et al. (2008a) and Graham \&{} Spitler (2009). In particular, Graham \&{} Spitler (2009) identify a dozen  NCs in galaxies that host SMBHs, whose mass was already determined via dynamical measurements. Furthermore, Graham \&{} Spitler (2009) suggest that most spheroids with stellar masses ranging from 10$^8 {\rm M}_\odot{}$ to a few $10^{10} {\rm M}_\odot{}$ might host both a SMBH and a NC.

NCs have been proposed as important sources of gravitational waves (GWs, Poincar\'e 1905; Einstein 1916, 1918), originating from mergers between stellar-mass black hole (BH) binaries (Miller \&{} Lauburg 2009; see also Freitag 2003; Hopman, Freitag \&{} Larson 2007; O'Leary, Kocsis \&{} Loeb 2009; O'Shaughnessy, Kalogera \&{} Belczynski 2010). In fact, because of their relatively high escape velocity, NCs should be able to retain most of their stellar-mass BHs (Miller \&{} Lauburg 2009), in spite of natal kicks and gravitational interactions (although the presence of a SMBH inside the NC might affect this scenario, see, e.g., Merritt 2006, 2009). Furthermore, three-body interactions are expected to be very frequent in NCs, given their high central density, leading to shrinking and coalescence of hard binaries. For these reasons, Miller \&{} Lauburg (2009) predict that tens of BH$-$BH mergers per year in NCs will be observable with the advanced Laser Interferometer Gravitational-wave Observatory (advanced LIGO, Harry et al. 2010). Given the increasing evidence of the coexistence between  SMBHs and  NCs in some galaxies, we propose that mergers between SMBHs and stellar BHs may also occur in NCs. Such events fall 
 within the class of
extreme mass-ratio inspirals (EMRIs, see Amaro-Seoane et al. 2007, for a review) and may be detected by space-borne interferometers.
 At present, it is still fairly uncertain what mission could fly in the future. As reference, we consider the published sensitivity performance of the Laser Interferometer Space Antenna (LISA, Bender et al. 1998; Gair, Tang \&{} Volonteri 2010), the European New Gravitational wave Observatory (NGO, derived from the previous LISA proposal, Amaro-Seoane et al. 2012a, 2012b) and the DECi-hertz Interferometer Gravitational wave Observatory (DECIGO, Kawamura et al. 2006). 

In this paper, we investigate the capture rate of BHs onto SMBHs in NCs which yield EMRIs\footnote{In the following, we will use the definition of EMRIs as merger events occurring from an orbit with semi-major axis, at the moment of capture by the SMBH (the capture radius is defined as $8\,{}r_g$, $r_g$ being the event horizon of the SMBH), less than $10^{-5}$ pc (Merritt et al. 2011).} and we estimate their detectability by future gravitational-wave space-based observatories.

\section{Method}
In this section, we derive the rate of EMRIs in NCs  and we infer an expected detection rate of such events by LISA, NGO and DECIGO.
\subsection{The detection rate}
The detection rate $R$ of EMRIs in NCs
can be expressed as (see, e.g., Miller 2002; Mapelli et al. 2010)
\begin{eqnarray}\label{eq:general}
R=\int_{m_1}^{m_2}{\rm d}m_{\rm SMBH}\,{}\times{} \nonumber\\
\int_0^{z_{\rm max}(m_{\rm SMBH}, m_{\rm co})}{\rm d}z\frac{{\rm d}^3N_{\rm EMRI}}{{\rm d}m_{\rm SMBH}\,{}{\rm d}t_{\rm e}\,{}{\rm d}V_{\rm c}}\,{}\frac{{\rm d}t_{\rm e}}{{\rm d}{t_{\rm o}}}\,{}\frac{{\rm d}V_{\rm c}}{{\rm d}z},
\end{eqnarray}
where $m_1$ and $m_2$ are the minimum and maximum SMBH mass ($m_{\rm SMBH}$); $N_{\rm EMRI}$ is the number of EMRIs of SMBH$-$BH binaries; $t_{\rm e}$ and $t_{\rm o}$ are the time in the rest frame of the source and of the observer, respectively (thus, $\frac{{\rm d}t_{\rm e}}{{\rm d}{t_{\rm o}}}=(1+z)^{-1}$). $V_c$ is the comoving volume. In 
 flat $\Lambda{}$ cold dark matter ($\Lambda{}$CDM), 
\begin{equation}\label{eq:volume}
\frac{{\rm d}V_{\rm c}}{{\rm d}z}=4\,{}\pi{}\,{}\left(\frac{c}{H_0}\right)^3\,{}\left(\int_0^z\frac{{\rm d}\tilde{z}}{{\mathcal E}(\tilde{z})}\right)^2\,{}\frac{1}{{\mathcal E}(z)}, 
\end{equation}
 where $c$ is the light speed, $H_0$ is the Hubble constant ($H_0=71$ km s$^{-1}$ Mpc$^{-1}$) and ${\mathcal E}(z)=\left[\Omega{}_\Lambda{}+(1+z)^3\Omega{}_{\rm M}\right]^{1/2}$, with $\Omega{}_\Lambda{}=0.73$ and $\Omega{}_{\rm M}=0.27$ (Larson et al. 2011).

The quantity $z_{\rm max}(m_{\rm SMBH}, m_{\rm co})$ is  the maximum redshift at which an event can be detected with a sky-location and orientation averaged signal-to-noise ratio $\langle{\mathrm SNR}\rangle \ge{}10$ by a single interferometer. Thus, $z_{\rm max}(m_{\rm SMBH}, m_{\rm co})$ defines the instrumental horizon of a given detector. In observations with a network of instruments, the signal-to-noise ratio scales as the square root of the number of instruments, and in this respect the results presented here should be considered as conservative.
The maximum redshift depends on (i) the mass of the SMBH $m_{\rm SMBH}$, (ii) the mass of the companion ($m_{\rm co}$) that merges with the SMBH, (iii) the orbital eccentricity $e(t)$, (iv) the spin parameter of the SMBH ($j=S\,{}c/G\,{}m_{\rm SMBH}^2$, where $S$ is the modulus of the angular momentum of the SMBH, $c$ is the speed of light and $G$ is the gravitational constant), and on various other parameters, as well as on the sensitivity of the instrument. In this paper, we adopt $m_{\rm co}=10\,{}{\rm M}_\odot{}$, corresponding to the typical mass of a stellar-mass BH. We consider different values of the eccentricity and of the spin parameter. See  Appendix~A for details on the derivation of $z_{\rm max}(m_{\rm SMBH}, m_{\rm co})$.

The EMRI rate per unit mass, time and volume can be written as:
\begin{equation}\label{eq:merg}
\frac{{\rm d}^3N_{\rm EMRI}}{{\rm d}m_{\rm SMBH}\,{}{\rm d}t_{\rm e}\,{}{\rm d}V_{\rm c}}=\nu{}_{\rm EMRI}\,{}\frac{{\rm d}n_{\rm SMBH}}{{\rm d}m_{\rm SMBH}},
\end{equation}
where $n_{\rm SMBH}$ is the comoving density of SMBHs surrounded by NCs and $\nu{}_{\rm EMRI}$ is the rate of EMRIs (i.e., the rate of mergers that end-up as EMRIs) per  SMBH.

Graham \&{} Spitler (2009) indicate that SMBHs coexist with NCs in spheroids with mass $10^8\le{}m_{\rm sph}/{\rm M}_\odot{}\le{}10^{10}$. The questions (i) whether there is a scaling between the mass of the SMBH and that of the host spheroid, and (ii) whether such relation is valid for all SMBH masses have been debated for a long time. Marconi \&{} Hunt (2003, hereafter MH03) find that $m_{\rm SMBH}\sim{}0.002\,{}m_{\rm sph}$, but there might be large deviations (up to a factor of 10) from this value for low-mass ($\le{}10^6\,{}$M$_\odot{}$) SMBHs (see, e.g., tab.~2 of Graham \&{} Spitler 2009).  Graham  (2012a, hereafter G12a) and Graham (2012b, hereafter G12b) propose that  the mass of the SMBH scales (nearly) with the square of the host spheroid mass in galaxies with NCs (i.e. galaxies with $m_{\rm sph}<10^{11}$ M$_\odot{}$), while it scales linearly with the  host spheroid mass in high-mass galaxies (which often have partially depleted cores). In particular, G12a proposes that
$\log_{10}(m_{\rm SMBH}/{\rm M}_{\odot{}}) = (8.38 \pm{} 0.17)\,{}+\,{}(1.92 \pm{} 0.38)\,{} \log_{10}(m_{\rm sph} / 7\times{}10^{10}{\rm M}_{\rm \odot{}})$ for galaxies with NCs. 
The BH mass range corresponding to $10^8\le{}m_{\rm sph}/{\rm M}_\odot{}\le{}10^{10}$ is $2\times{}10^5-2\times{}10^7$ M$_{\odot{}}$ and $8\times{}10^2-5.7\times{}10^6$ M$_{\odot{}}$, according to the results by  MH03 and by G12a, respectively. In the following, we will consider both cases. In particular, we define 
$f_{\rm BH}(m_{\rm sph})= 0.002\,{}m_{\rm sph}$ and $f_{\rm BH}(m_{\rm sph})= 10^{8.38}{\rm M}_\odot{}\,{}\left(m_{\rm sph}/7\times{}10^{10}{\rm M}_\odot{}\right)^{1.92}$, when adopting the formalism by   MH03 and by  G12a, respectively. 
Therefore, equation~(\ref{eq:general}) can be rewritten as
\begin{eqnarray}\label{eq:sph}
R=4\,{}\pi{}\,{}\left(\frac{c}{H_0}\right)^3\int_{m_{\rm sph1}}^{m_{\rm sph2}}\,{}{\rm d}m_{\rm sph}\,{}\nu{}_{\rm EMRI}\times{}\nonumber{}\\
\int_0^{z_{\rm max}[f_{\rm BH}(m_{\rm sph}), m_{\rm co}]}{\rm d}z\,{}f_{\rm bb}\,{}\frac{{\rm d}n_{\rm sph}}{{\rm d}m_{\rm sph}}\times{}\nonumber{}\\
\,{}\frac{1}{(1+z)\,{}{\mathcal E}(z)}\,{}\left(\int_0^z \frac{{\rm d}\tilde{z}}{{\mathcal E}(\tilde{z})}\right)^2,
\end{eqnarray}
where $m_{\rm sph1}=10^8\,{}{\rm M}_\odot{}$ and $m_{\rm sph2}=10^{10}\,{}{\rm M}_\odot{}$ are the minimum and the maximum mass of spheroids that host SMBHs together with NCs, respectively; $f_{\rm bb}$ is the fraction of spheroids in the considered mass range ($10^8\le{}m_{\rm sph}/{\rm M}_\odot{}\le{}10^{10}$) that host both a SMBH and a NC (we assume $f_{\rm bb}=1$, according to Graham \&{} Spitler 2009); $n_{\rm sph}$ is the comoving density of spheroids. In equation~\ref{eq:sph}, we used the fact that $\frac{{\rm d}n_{\rm SMBH}}{{\rm d}m_{\rm SMBH}}=f_{\rm bb}\,{}\frac{{\rm d}m_{\rm sph}}{{\rm d}m_{\rm SMBH}}\,{}\frac{{\rm d}n_{\rm sph}}{{\rm d}m_{\rm sph}}$.

We can derive $n_{\rm sph}$ from the comoving density of halos $n_{\rm h}$, assuming that spheroids with mass $m_{\rm sph}$ are located in halos with total mass $m_{\rm h}=m_{\rm sph}/p$. Thus, equation~(\ref{eq:sph}) can be rewritten as

\begin{eqnarray}\label{eq:halo}
R=4\,{}\pi{}\,{}\left(\frac{c}{H_0}\right)^3\int_{m_{\rm h1}}^{m_{\rm h2}}\,{}{\rm d}m_{\rm h}\,{}\nu{}_{\rm EMRI}\times{}\nonumber{}\\
\int_0^{z_{\rm max}[f_{\rm BH}(p\,{}m_{\rm h}), m_{\rm co}]}{\rm d}z\,{}\,{}f_{\rm bb}\,{}\frac{{\rm d}n_{\rm h}}{{\rm d}m_{\rm h}}\times{}\nonumber{}\\
\frac{1}{(1+z)\,{}{\mathcal E}(z)}\,{}\left(\int_0^z \frac{{\rm d}\tilde{z}}{{\mathcal E}(\tilde{z})}\right)^2,
\end{eqnarray}
where  $\frac{{\rm d}n_{\rm h}}{{\rm d}m_{\rm h}}$ is the Press-Schechter function (Press \&{} Schechter 1974; Eisenstein \&{} Hu 1998, 1999),  $m_{\rm h1}=m_{\rm sph1}/p$ and $m_{\rm h2}=m_{\rm sph2}/p$.
We use $p=1.06\times{}10^{-2}$ as a fiducial value, but we consider also  $p=5.04\times{}10^{-2}$ and  $=2.81\times{}10^{-3}$ as upper and lower values, respectively, depending on the baryonic fraction in  dark matter halos (see Appendix~B for the derivation of $p$). In the equation~(\ref{eq:halo}), we write $f_{\rm bb}$ inside both integrals (in mass and redshift), although, in the following, we will assume for simplicity that $f_{\rm bb}$ is a constant.

\subsection{The EMRI rate}
The EMRI rate $\nu{}_{\rm EMRI}$ (i.e. the rate of mergers that end-up as EMRIs) is likely the most difficult quantity to estimate in equation~(\ref{eq:halo}), as it depends on the dynamics and on the relativistic effects in the neighborhoods ($\lesssim{}10^{-2}$ pc) of the SMBH. In particular, the collisional dynamics of relativistic star clusters around SMBHs is poorly understood. The presence of the SMBH potential well strongly affects the inner regions of the star cluster and the dynamics of binary stars (Alexander 1999, 2003, 2005; Alexander \&{} Hopman 2009; Hopman 2009). To derive $\nu{}_{\rm EMRI}$ in an accurate way is beyond the aims of this paper. 
Here, we will adopt approximate rates derived by Merritt et al. (2011).
Capture of stars on EMRI orbits can be driven by resonant relaxation (Rauch \&{} Tremaine 1996; Hopman \&{} Alexander 2006a), by two-body relaxation (Merritt et al. 2011) or even by dynamical friction (Antonini \&{} Merritt 2012),  that is by physical processes where background stars exert torques on the stellar orbits. The capture of a star is hampered by the so-called Schwarzschild barrier (Merritt et al. 2011), that is by an angular-momentum barrier associated with the value of the orbital angular momentum at which the background torques become ineffective due to the Schwarzschild precession of the orbit\footnote{Background torques are effective only on a timescale shorter than the fastest mechanism that changes the relative orientation of a star with respect to the gravitational field of the background stars. It can be shown that, for high eccentricity, Schwarzschild precession is the fastest mechanism (Merritt et al. 2011).}. Merritt et al. (2011) analyze various mechanisms to penetrate the Schwarzschild barrier (based on classical non-resonant relaxation), and derive a time scale $t_{\rm loss}$ for stars to `definitely' cross the barrier (see their equation 71 and tables 2-3).

According to the results by Merritt et al. (2011), we approximate  $\nu{}_{\rm EMRI}$ as:
\begin{equation}\label{eq:eqmerg}
\nu{}_{\rm EMRI}=10^{-7}\,{}{\rm yr}^{-1}\left(\frac{\tilde{N}}{1}\right)\,{}\left(\frac{t_{\rm loss}}{10^7\,{}{\rm yr}}\right)^{-1},
\end{equation}
where $\tilde{N}$ is the number of stars initially with semi-major axis (with respect to the SMBH) $a\sim{}2-10\times{}10^{-3}$ pc. $\tilde{N}$ is basically unknown. We also note that the adopted value of $t_{\rm loss}$ was derived for a SMBH-to-BH mass ratio equal to $2\times{}10^4$. Therefore it is suited for $m_{\rm SMBH}=2\times{}10^5$ M$_\odot{}$ (adopting $m_{\rm co}=10$ M$_\odot{}$). On the other hand, most of the contribution for the detection rates (in Table~1) comes from such `low-mass' SMBHs, as we will discuss in the next section.

We stress that, at the moment of capture, the eccentricity is likely very high (e.g., Merritt et al. 2011, and references therein; but see Miller et al. 2005 for the effect of tidal disruption of binaries), and GWs are emitted in short pulses during pericentre passages. Then, the orbit gradually shrinks and circularizes (over a timescale of $\sim{}10^3-10^8$ yr, e.g. Barack \&{} Cutler 2004b), because of GW emission, and the GW emission can be observed as continuous. In the next Section, we define $e_{\rm LSO}$ as the eccentricity at the last stable orbit (LSO) and we integrate it back in time accounting for GW circularization (see the Appendix A and Barack \&{} Cutler 2004a).

 We note also that Merritt et al. (2011) adopt a steady-state, cuspy model of the galactic centre (e.g.,  Sigurdsson \&{} Rees 1997; Hopman \&{} Alexander 2006b; Freitag, Amaro-Seoane \&{} Kalogera 2006), to derive the estimate that we use in equation~(\ref{eq:eqmerg}). In these models, which are relaxed and strongly mass segregated, the mass within $\sim{}0.1$ pc of the SMBH is contributed mainly by stellar-mass BHs. Therefore, two-body scatterings dominate the orbital evolution in proximity of the Schwarzschild barrier (e.g., Freitag et al. 2008; Alexander \&{} Hopman 2009), whereas dynamical friction is negligible.

If stars in the central parsec of a galaxy follow a flat core rather than a relaxed cusp, the time for a stellar-mass BH (initially located out of the core) to reach the centre of the host galaxy can easily exceed the Hubble time (Merritt 2010). 
Thus, in flat cores, the number of stellar BHs within the influence radius of the SMBH is still low  with respect to the total number of objects, including lower-mass remnants and stars. In this case, dynamical friction can be more important than two-body scatterings to penetrate the Schwarzschild barrier  (according to the new, more general expression of dynamical friction, derived by Antonini \&{} Merritt 2012). This likely leads to lower EMRI rates in the case of flat cores and deserves further studies.

Whether the galaxies hosting a NC have a central stellar density cusp or a flat core is an open question. The MW, which was long believed to have a steeply-rising mass density near the SMBH, was recently found to have a parsec-scale core in the old red giant branch (RGB) stellar component (Buchholz, Sch\"odel \&{} Eckart 2009; Do et al. 2009; Bartko et al. 2010). Other MW-like galaxies, hosting both a NC and a SMBH at their centre, might behave in the same way. Therefore, steady-state mass segregated models might be inappropriate to describe a MW-like galactic centre.
On the other hand, only the RGB stars were observed to follow a flat core distribution in the central parsec of the MW, while the non-relaxed  young stars show a rather cuspy profile (e.g., Do et al. 2009). We do not know whether the old main-sequence stars and  especially  the stellar remnants follow the same cored distribution as RGB stars (e.g., Dale et al. 2009; Davies et al. 2011). 

 A further possible caveat is that a currently cored profile might have been cuspy in the past, and then the cusp was destroyed as a consequence of dynamical effects (e.g., Merritt 2010; Yusef-Zadeh, Bushouse \&{} Wardle 2012). 
At the moment, the most likely scenario to explain the removal of the cusp and its transformation into a core is that of binary scouring as a consequence of a galaxy merger (Gualandris \&{} Merritt 2012). In this case, the central cusp of stellar BHs reforms in a time that is at least one relaxation time, and the number of BHs in the proximity of the SMBH is smaller than in relaxed multi-mass models.

In summary,  whether galaxies hosting NCs have a central stellar cusp or a core, and for how long these profiles survive are very uncertain issues. Our predictions for the detectable EMRI rate, based on a mass-segregated cuspy model, represent the most optimistic case and should be regarded as upper limits.

\section{Results}
\subsection{Maximum redshift}

\begin{figure}
  \resizebox{7cm}{!}{\includegraphics{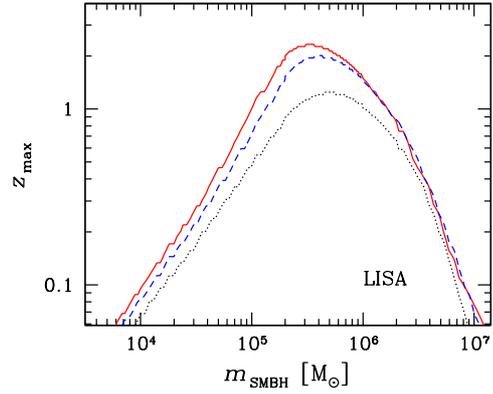}}
  \caption{$z_{\rm max}$ in the case of LISA, as a function of $m_{\rm SMBH}$ for three different simulations. In all the shown models, $j=1$ and  $T_{\rm mission}=t_0=5$ yr (see Table~1). Solid line (red on the web): model with $e_{\rm LSO}=0.3$. Dashed line (blue on the web): model with $e_{\rm LSO}=0.5$. Dotted black line: model with $e_{\rm LSO}=0.7$.}
  \label{fig:fig1}
\end{figure}
\begin{figure}
  \resizebox{7cm}{!}{\includegraphics{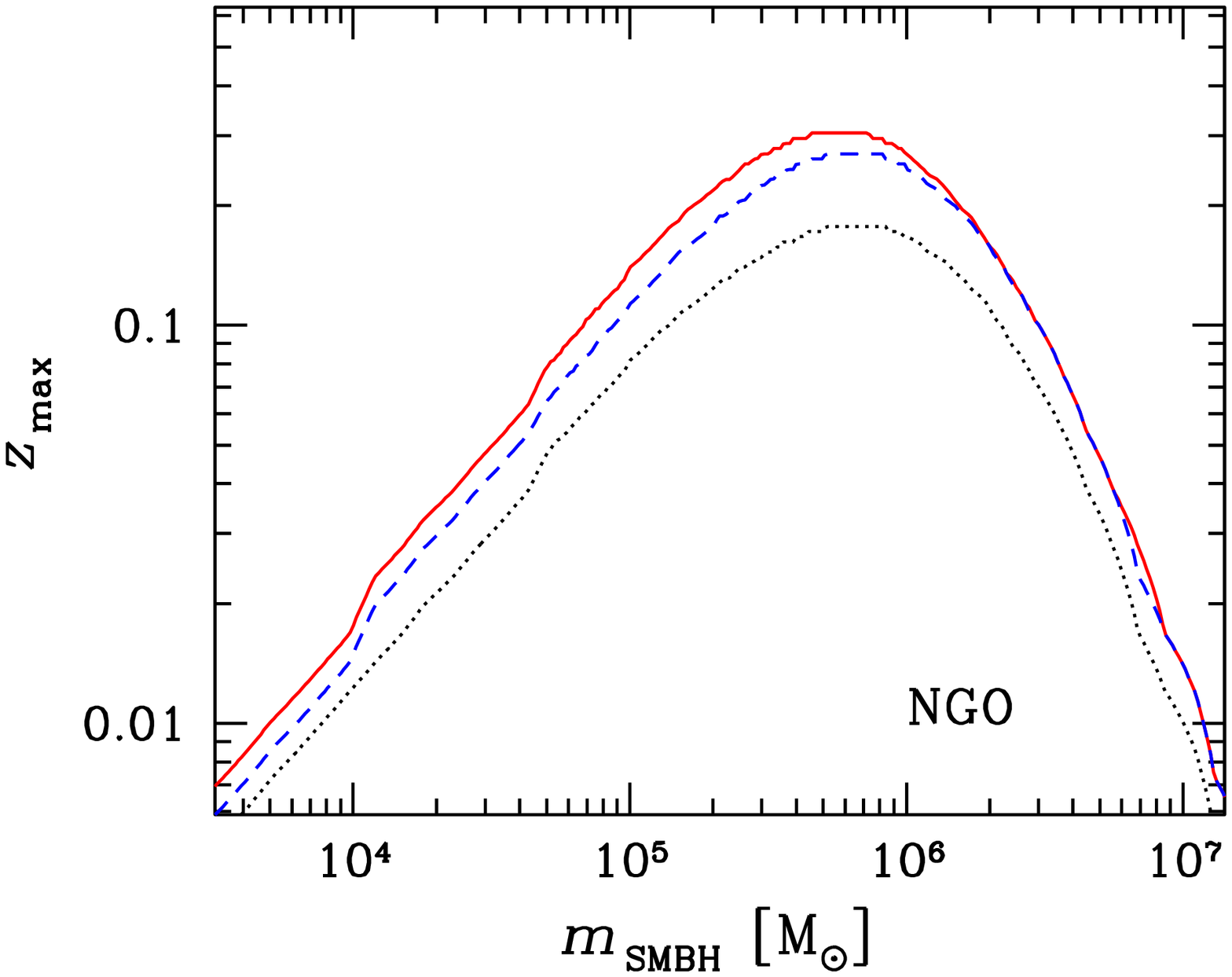}}
  \caption{$z_{\rm max}$ in the case of NGO, as a function of $m_{\rm SMBH}$ for three different simulations. In all the shown models, $j=1$ and  $T_{\rm mission}=t_0=5$ yr (see Table~1). Solid line (red on the web): model with $e_{\rm LSO}=0.3$. Dashed line (blue on the web): model with $e_{\rm LSO}=0.5$. Dotted black line: model with $e_{\rm LSO}=0.7$. 
}
  \label{fig:fig2}
\end{figure}
\begin{figure}
  \resizebox{7cm}{!}{\includegraphics{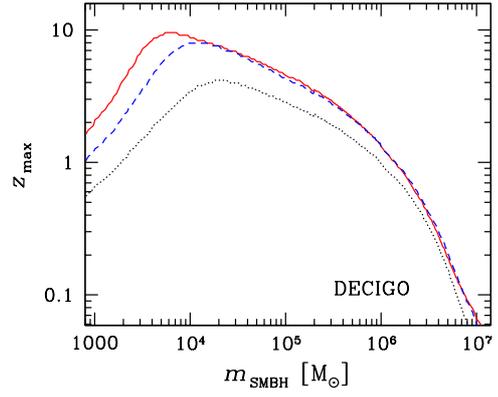}}
  \caption{$z_{\rm max}$ in the case of DECIGO, as a function of $m_{\rm SMBH}$ for three different simulations. In all the shown models, $j=1$ and  $T_{\rm mission}=t_0=5$ yr (see Table~1). Solid line (red on the web): model with $e_{\rm LSO}=0.3$. Dashed line (blue on the web): model with $e_{\rm LSO}=0.5$. Dotted black line: model with $e_{\rm LSO}=0.7$. 
}
  \label{fig:fig3}
\end{figure}

 Figs.~\ref{fig:fig1}, ~\ref{fig:fig2}  and ~\ref{fig:fig3} show the maximum redshift ($z_{\rm max}$) as a function of $m_{\rm SMBH}$ in the case of LISA, NGO and DECIGO, respectively. In these Figures, $t_0=5$ yr and $T_{\rm mission}=5$ yr. Here, the value $t_0$ is the assumed time before the plunge-in, i.e. the time elapsed between the beginning of the observation with the gravitational interferometer and the moment in which the companion reaches the LSO (in this paper, we study the cases where $t_0=5,\,{}6,\,{}10$ and 15 yr). $T_{\rm mission}$ is defined as the duration of the mission. In this paper, we consider $T_{\rm mission}=5$ and $2$ yr as two fiducial values of the mission duration, for the three considered gravitational experiments (see, e.g., Gair et al. 2010).  We remind that $z_{\rm max}$ was obtained for a sky-location and orientation averaged signal-to-noise ratio higher than 10 (see Section 2.1 and the Appendix~A).

In the case of LISA (Fig.~\ref{fig:fig1}), $z_{\rm max}$  reaches high values ($z\sim{}2$) for $m_{\rm SMBH}\sim{}10^5-10^6{\rm M}_\odot{}$ and eccentricity at the LSO $e_{\rm LSO}\le{}0.5$.  The value of $z_{\rm max}$ for LISA drops below $\sim{}1$ for   $m_{\rm SMBH}<10^5{\rm M}_\odot{}$ and for  $m_{\rm SMBH}>2\times{}10^6{\rm M}_\odot{}$.
The behaviour of $z_{\rm max}$ for NGO (Fig.~\ref{fig:fig2}) is similar to that of LISA, in the sense that  $z_{\rm max}$ drops quite fast for  $m_{\rm SMBH}<10^5{\rm M}_\odot{}$ and for  $m_{\rm SMBH}>2\times{}10^6{\rm M}_\odot{}$. On the other hand, the maximum value of $z_{\rm max}$ for NGO (obtained for $m_{\rm SMBH}\sim{}10^5-10^6{\rm M}_\odot{}$) is $z_{\rm max}\sim{}0.3$, much lower than in the case of LISA.

 In the case of DECIGO (Fig.~\ref{fig:fig3}), $z_{\rm max}\sim{}3$ for $m_{\rm SMBH}\sim{}10^5{\rm M}_\odot{}$, but $z_{\rm max}$ keeps rising for smaller BH masses and reaches a value of $8-10$ for $m_{\rm SMBH}\sim{}5\times{}10^3{\rm M}_\odot{}$. We stress that the existence of BHs in the $10^3-10^5$ M$_\odot{}$ mass range is still controversial (e.g., Miller \&{} Colbert 2004, and references therein).

\begin{table}
\caption{Simulated values of the rate $R$ for the old LISA configuration.}
\centering
\begin{tabular}[!h]{llllll}
\hline\hline
$j$
& $e_{\rm LSO}$
& $t_0$
& $T_{\rm mission}$ 
& $p/10^{-2}$ 
& $R$ [yr$^{-1}$]\\
\hline
0.1&0.1& 5 yr & 5 yr & 1.06 & 400 (4700)\\ 
0.1&0.3& 5 yr & 5 yr & 1.06 & 300 (4500)\\
0.1&0.5& 5 yr & 5 yr & 1.06 & 200 (3100)\\
0.1&0.7& 5 yr & 5 yr & 1.06 &  80 (1200)\\
\vspace{0.1cm}\\
0.5&0.1& 5 yr & 5 yr & 1.06 & 500 (4700)\\
0.5&0.3& 5 yr & 5 yr & 1.06 & 300 (4700)\\
0.5&0.5& 5 yr & 5 yr & 1.06 & 200 (3600)\\
0.5&0.7& 5 yr & 5 yr & 1.06 & 100 (1500)\\
\vspace{0.1cm}\\
1&0.1& 5 yr & 5 yr & 1.06  &  500  (4900)\\
1&0.1& 5 yr & 5 yr & 5.04  & 2200 (19900)\\
1&0.1& 5 yr & 5 yr & 0.281 &  100  (1400)\\
1&0.3& 5 yr & 5 yr & 1.06  &  400  (5300)\\
1&0.5& 5 yr & 5 yr & 1.06  &  300  (4300)\\
1&0.7& 5 yr & 5 yr & 1.06  &  100  (1900)\\
\vspace{0.1cm}\\
1&0.5& 6 yr & 5 yr & 1.06 &  100  (1700)\\
1&0.5& 10 yr & 5 yr & 1.06 &  20   (300)\\
1&0.5& 15 yr & 5 yr & 1.06 &   8   (100)\\
\vspace{0.1cm}\\
1&0.5& 5 yr & 2 yr & 1.06 &   20   (300)\\
1&0.5& 6 yr & 2 yr & 1.06 &   10   (200)\\
\noalign{\vspace{0.1cm}}
\hline
\end{tabular}
\tablefoot{The tabulated values of $R$ depend on the spin parameter ($j$), on the eccentricity at the last stable orbit ($e_{\rm LSO}$), on the initial time of observations, defined as time before the plunge-in ($t_0$), on the expected duration of the mission ($T_{\rm mission}$), on  $p\equiv{}m_{\rm sph}/m_{\rm h}$ (see Section 2.1 and Appendix~B), and on $f_{\rm BH}$ (i.e. the adopted $m_{\rm SMBH}-m_{\rm sph}$ relation). The value of $R$ out of (inside) parentheses refers to $f_{\rm BH}$ from G12a (MH03). Tabulated rates are for $\tilde{N}=1$ and $t_{\rm loss}=10^7$ yr.}
\end{table}
\begin{table}
\caption{The same as Table 1, but for the NGO configuration.}
\centering
\begin{tabular}[!h]{llllll}
\hline\hline
$j$
& $e_{\rm LSO}$
& $t_0$
& $T_{\rm mission}$\tablefootmark{a}
& $p/10^{-2}$ 
& $R$ [yr$^{-1}$]\\
\hline
0.1&0.1& 5 yr & 5 yr & 1.06 & 4 (50)\\ 
0.1&0.3& 5 yr & 5 yr & 1.06 & 3 (40)\\
0.1&0.5& 5 yr & 5 yr & 1.06 & 2 (20)\\
0.1&0.7& 5 yr & 5 yr & 1.06 & 0.6 (7)\\
\vspace{0.1cm}\\
0.5&0.1& 5 yr & 5 yr & 1.06 & 5 (50)\\
0.5&0.3& 5 yr & 5 yr & 1.06 & 4 (50)\\
0.5&0.5& 5 yr & 5 yr & 1.06 & 2 (30)\\
0.5&0.7& 5 yr & 5 yr & 1.06 & 0.7 (9)\\
\vspace{0.1cm}\\
1&0.1& 5 yr & 5 yr & 1.06  &   6  (70)\\
1&0.1& 5 yr & 5 yr & 5.04  &   20 (250)\\
1&0.1& 5 yr & 5 yr & 0.281 &   2 (20)\\
1&0.3& 5 yr & 5 yr & 1.06  &   5 (60)\\
1&0.5& 5 yr & 5 yr & 1.06  &   3  (40)\\
1&0.7& 5 yr & 5 yr & 1.06  &   1  (10)\\
\vspace{0.1cm}\\
1&0.5& 6 yr & 5 yr & 1.06 &   0.6 (10)\\
1&0.5& 10 yr & 5 yr & 1.06 &  0.1 (1)\\
1&0.5& 15 yr & 5 yr & 1.06 &  0.01 (0.2)\\
\vspace{0.1cm}\\
1&0.5& 5 yr & 2 yr & 1.06 &   0.1 (1)\\
1&0.5& 6 yr & 2 yr & 1.06 &   0.07 (0.8)\\
\noalign{\vspace{0.1cm}}
\hline
\end{tabular}
\tablefoottext{a}{Nominal duration of the NGO mission is 2 yr.}
\end{table}
\begin{table}
\caption{The same as Table 1, but for DECIGO.}
\centering
\begin{tabular}[!h]{llllll}
\hline\hline
$j$
& $e_{\rm LSO}$
& $t_0$
& $T_{\rm mission}$
& $p/10^{-2}$
& $R$ [yr$^{-1}$]\\
\hline
0.1&0.1& 5 yr & 5 yr & 1.06 & 11400 (5900)\\ 
0.1&0.3& 5 yr & 5 yr & 1.06 & 6800 (6400)\\
0.1&0.5& 5 yr & 5 yr & 1.06 & 4700 (5900)\\
0.1&0.7& 5 yr & 5 yr & 1.06 & 2100 (3300)\\
\vspace{0.1cm}\\
0.5&0.1& 5 yr & 5 yr & 1.06 & 11500 (5900)\\
0.5&0.3& 5 yr & 5 yr & 1.06 & 8100  (6400)\\
0.5&0.5& 5 yr & 5 yr & 1.06 & 5900  (6000)\\
0.5&0.7& 5 yr & 5 yr & 1.06 & 3000  (3600)\\
\vspace{0.1cm}\\
1&0.1& 5 yr & 5 yr & 1.06 &  11600 (5600)\\
1&0.1& 5 yr & 5 yr & 5.04 & 56100 (23500)\\
1&0.1& 5 yr & 5 yr & 0.281 &  2900 (1600)\\
1&0.3& 5 yr & 5 yr & 1.06 &   9800 (6400)\\
1&0.5& 5 yr & 5 yr & 1.06 &   7700 (6300)\\
1&0.7& 5 yr & 5 yr & 1.06 &   4600 (4000)\\
\vspace{0.1cm}\\
1&0.5& 6 yr & 5 yr & 1.06 &   100 (2000)\\
1&0.5& 10 yr & 5 yr & 1.06 &  20 (200)\\
1&0.5& 15 yr & 5 yr & 1.06 &  4 (50)\\
\vspace{0.1cm}\\
1&0.5& 5 yr & 2 yr & 1.06 &  30  (400)\\
1&0.5& 6 yr & 2 yr & 1.06 &  20 (200)\\
\noalign{\vspace{0.1cm}}
\hline
\end{tabular}
\footnotesize{}
\end{table}

\begin{figure}
  \resizebox{7cm}{!}{\includegraphics{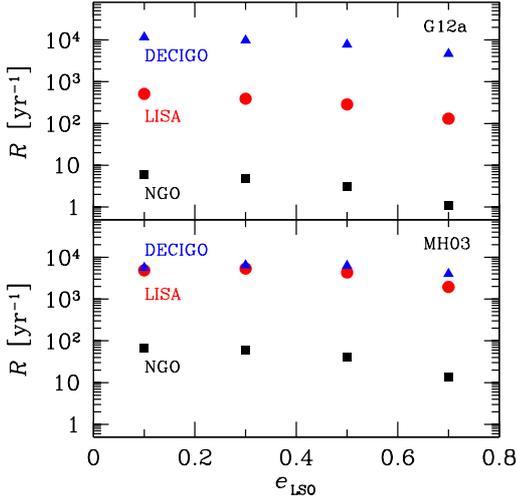}}
  \caption{ The total detection rate $R$ versus the adopted eccentricity at the LSO ($e_{\rm LSO}$), in the case of LISA (filled circles, red on the web), DECIGO (filled triangles, blue on the web) and NGO (filled squares). For each case, $j=1$, $t_0=5$ yr, $T_{\rm mission}=5$ yr (see Tables~1, 2 and 3), $p=1.06\times{}10^{-2}$, $\tilde{N}=1$ and $t_{\rm loss}=10^7$ yr. Top panel: $m_{\rm SMBH}-m_{\rm sph}$ relation set according to G12a. Bottom panel:  $m_{\rm SMBH}-m_{\rm sph}$ relation set according to MH03.
}
  \label{fig:fig4}
\end{figure}
\begin{figure}
  \resizebox{7cm}{!}{\includegraphics{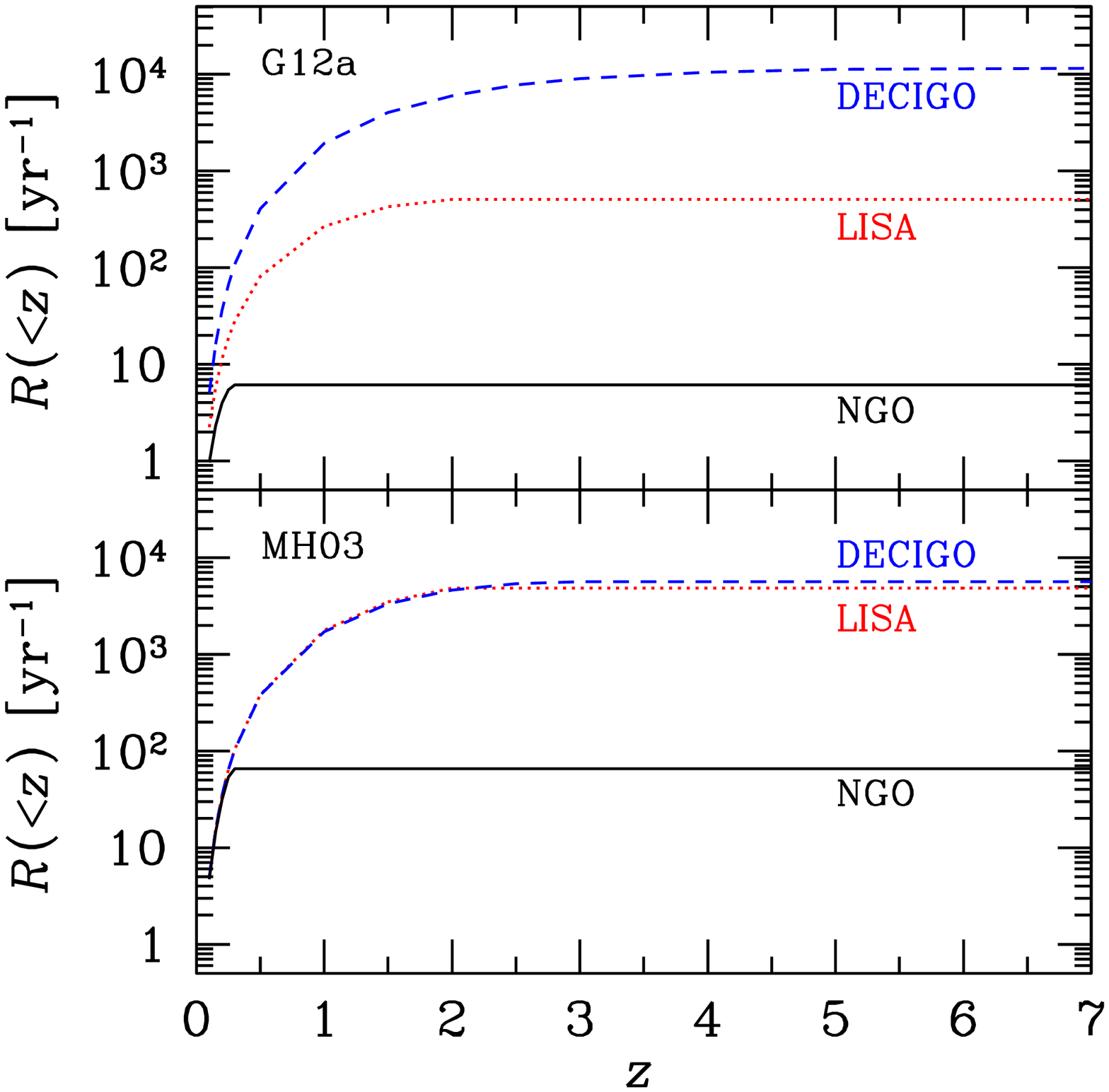}}
  \caption{ Cumulative redshift distribution of the expected detections per year, in the case of LISA (dotted line, red on the web), DECIGO (dashed line, blue on the web) and NGO (solid line). For each case, $j=1$, $e_{\rm LSO}=0.1$, $t_0=5$ yr, $T_{\rm mission}=5$ yr (see Tables~1, 2 and 3), $p=1.06\times{}10^{-2}$, $\tilde{N}=1$ and $t_{\rm loss}=10^7$ yr. Top panel: $m_{\rm SMBH}-m_{\rm sph}$ relation set according to G12a. Bottom panel:  $m_{\rm SMBH}-m_{\rm sph}$ relation set according to MH03. 
}
  \label{fig:fig5}
\end{figure}

\subsection{Detection rates}
 Tables~1, 2 and 3 summarize the estimated values of the detection rate $R$ for the simulated models, in the case of LISA, NGO and DECIGO, respectively. 

Fig.~\ref{fig:fig4} shows the behaviour of $R$ as a function of $e_{\rm LSO}$ in the case of LISA, NGO and DECIGO. In the top panel of Fig.~\ref{fig:fig4}, the  $m_{\rm SMBH}-m_{\rm sph}$ relation follows the prescriptions by G12a, while in the bottom panel it follows the linear scaling derived by MH03.
Tables~1, 2, 3 and Fig.~\ref{fig:fig4} indicate that LISA and DECIGO are expected to detect a factor of $\gtrsim{}100$ more EMRI events than NGO.

A comparison among Tables~1, 2 and 3 indicates that LISA, NGO and DECIGO have similar responses to the physical parameters influencing EMRI events (although the overall sensitivity of NGO is much lower than that of the other two instruments). In particular, for all the considered space-borne instruments, $R$ is slightly affected by the spin parameter.
 $R$ mildly depends on the eccentricity (Fig.~\ref{fig:fig4}): values of $e_{\rm LSO}\ge{}0.7$ often correspond to lower values of $R$, since $z_{\rm max}$ is lower for $e_{\rm LSO}\ge{}0.7$.

Tables~1, 2 and 3 show that the detection rate is very sensitive to the parameter $t_0$, which indicates how far the observed systems are from final coalescence. $R$ rapidly decreases as $t_0$ increases. To accurately derive $R$, we should `average' our results over a distribution of values of  $t_0$. This is beyond the aims of this paper. Therefore, we will consider the estimates of $R$ obtained for $t_0=5$ yr as upper limits, as assuming $t_0=5$ yr maximizes the probability of observing a merger event for all the duration of the mission.
Furthermore, the detection rate sensibly depends on the duration of the mission ($T_{\rm mission}$). If we assume that LISA, NGO and DECIGO will operate only for two years and that $t_0=5$ yr, the expected rates are lower by a factor of $\approx{}10$. 
 
In our approximate model, $R$  scales linearly with $\nu_{\rm EMRI}$: this quantity is highly uncertain, but our estimates can be rescaled for different choices of $\nu_{\rm EMRI}$ (the adopted values can be considered as upper limits). 
Finally, the halo-to-spheroid mass ratio ($p$) changes dramatically the rate $R$, and is another poorly constrained factor.

From Fig.~\ref{fig:fig4} and from Tables~1, 2 and 3, we note also that there is a strong dependence of $R$ on the assumed model for $f_{\rm BH}(m_{\rm sph})$, that is on the assumed $m_{\rm SMBH}-m_{\rm sph}$ relation. In the case of both LISA and NGO, the relation derived by G12a predicts a factor of $\approx{}10$ less detections than the linear scaling obtained by MH03. This is not true for DECIGO, whose expected detection rate is almost unaffected by the assumed $f_{\rm BH}(m_{\rm sph})$.

The reason of this behaviour is quite intuitive: the  $m_{\rm SMBH}-m_{\rm sph}$ scaling relation obtained by G12a ($m_{\rm SMBH}\propto{}m_{\rm sph}^{1.9}$) implies that relatively low mass BHs ($<10^5$ M$_\odot{}$) are associated with most of the spheroids in the $10^8-10^{10}{\rm M}_\odot{}$ mass range, whereas the linear scaling relation derived by MH03 predicts that SMBHs with $2\times{}10^5{\rm M}_\odot{}\le{}m_{\rm SMBH}\le{}2\times{}10^7{\rm M}_\odot{}$ are hosted by spheroids in the same mass range as above. LISA and NGO have very low sensitivity to EMRIs with $m_{\rm SMBH}<10^5$ M$_\odot{}$ (Figs.~\ref{fig:fig1} and \ref{fig:fig2}), and this explains the low values of $R$ when the relation by  G12a is assumed.
Instead, DECIGO is particularly suitable to observe EMRIs involving $10^4$ M$_\odot{}$ BHs (Fig.~\ref{fig:fig3}), while performing as well as LISA  for  $m_{\rm SMBH}\sim{}10^5-10^6$ M$_\odot{}$. This explains why the predicted detection rate for DECIGO is almost unaffected by the adopted $m_{\rm SMBH}-m_{\rm sph}$ scaling relation.

 Fig.~\ref{fig:fig5} shows the cumulative redshift distribution of the expected detections per year, that is it shows how many detections per year are expected for EMRIs at redshift $<z$, as a function of $z$. Fig.~\ref{fig:fig5} confirms that NGO is blind for EMRIs with $z>0.3$ (as a consequence of the behaviour of $z_{\rm max}$, see Fig.~\ref{fig:fig2}). From Fig.~\ref{fig:fig5} we note also the very different behaviour of LISA and DECIGO with respect to the adopted $m_{\rm SMBH}-m_{\rm sph}$ scaling relation. When the MH03 relation is assumed, LISA and DECIGO have almost the same behaviour, in terms of both total detection rate and dependence on redshift: most of detectable EMRIs occur at $z\lesssim{}2$.
 Instead, when the G12a relation is assumed, most of detectable EMRIs occur at $z\lesssim{}2$ and  $z\lesssim{}4$ for LISA and DECIGO, respectively. 
We stress that our predictions based on the Press-Schechter formalism might overlook many important processes affecting BH and galaxy evolution at $z>>1$.

\subsection{Comparison with previous studies}
 In this section, we compare the rates derived for the original LISA configuration and for NGO with those found in previous studies (e.g., Gair et al. 2004; Rubbo, Holley-Bockelmann \&{} Finn 2006; Gair et al. 2010; Amaro-Seoane et al. 2012a). In particular, we will focus on  Gair et al. (2004) and on Amaro-Seoane et al. (2012a), for LISA and NGO, respectively.

 Gair et al. (2004) find a EMRI detection rate $\sim{}30-350$ yr$^{-1}$ (for a 10 M$_\odot{}$ BH), depending on the assumed properties of the LISA mission. This result is quite consistent with our predictions, when we assume the G12a model, and is a factor of $\sim{}10$ lower than our results, when we adopt the MH03 prescriptions. On the other hand, there are significant differences between our method and that by  Gair et al. (2004).

Firstly,  Gair et al. (2004) assume a EMRI rate $\nu{}_{\rm EMRI}\sim{}10^{-6}(m_{\rm SMBH}/3\times{}10^6)^{3/8}$ yr$^{-1}$ per SMBH (for mergers with a 10 M$_\odot{}$ BH). This is a factor of 10 higher than our assumed EMRI rate (see equation~\ref{eq:eqmerg}), when we adopt $\tilde{N}=1$ (as in our Figs.~\ref{fig:fig4} and ~\ref{fig:fig5}).

Secondly,  Gair et al. (2004) assume  a different mass function for the SMBHs with respect to us. In particular,   Gair et al. (2004) estimate $\frac{{\rm d}n_{\rm SMBH}}{{\rm d}m_{\rm SMBH}}$ from the $m_{\rm SMBH}-\sigma$ relation (where $\sigma{}$ is the velocity dispersion, e.g., Ferrarese \&{} Merritt 2000; Gebhardt et al. 2000; Graham et al. 2011) and from the luminosity versus  $\sigma$ relation (Aller \&{} Richstone 2002). Instead, we use the $m_{\rm SMBH}-m_{\rm sph}$ relation (e.g., G12a) and the Press-Schechter formalism (Press \&{} Schechter 1974). 
Furthermore, we consider only those SMBHs that can coexist with a NC (otherwise the capture rate would likely be much lower), whereas   Gair et al. (2004) take into account all the BHs in the mass range $10^5-5\times{}10^6$ M$_\odot{}$, excluding only those that are located in Sc and Sd galaxies.

These differences are important, in terms of EMRI rates, as it can be shown by calculating the expected merger rate per unit volume at redshift $z=0$ ($\tilde{\nu}_{\rm EMRI}\equiv{}n_{\rm SMBH}(z=0)\,{}\,{}\nu_{\rm EMRI}$, where $n_{\rm SMBH}(z=0)$ is the number density of SMBHs at $z=0$ for the considered mass range and $\nu_{\rm EMRI}$ is the EMRI rate per SMBH). In our model, considering only SMBHs with  $10^5\le{}m_{\rm SMBH}\le{}5\times{}10^6$ M$_\odot{}$  (i.e., the mass range adopted by  Gair et al. 2004), $\tilde{\nu}_{\rm EMRI}\approx{}20$ Gpc$^{-3}$ yr$^{-1}$
 and $\tilde{\nu}_{\rm EMRI}\approx{}2$ Gpc$^{-3}$ yr$^{-1}$,  assuming the MH03 and the G12a scaling relation, respectively.

 Our estimate of $\tilde{\nu}_{\rm EMRI}$ derived from the G12a  scaling relation is similar to the one reported in table~2 of  Gair et al. (2004), while it is one order of magnitude higher than the one reported by  Gair et al. (2004) when the MH03 scaling relation is adopted instead. Our estimate of $\tilde{\nu}_{\rm EMRI}$ based on G12a, although very similar to the one derived by   Gair et al. (2004), comes from the product between our merger rate ${\nu}_{\rm EMRI}$, which is a factor of 10 lower than the one adopted by   Gair et al. (2004), and the number density of SMBHs with $10^5\le{}m_{\rm SMBH}\le{}5\times{}10^6$ M$_\odot{}$, which is a factor of 10 higher than the one in  Gair et al. (2004). We stress that our value of $\tilde{\nu}_{\rm EMRI}$ is an upper limit, as  ${\nu}_{\rm EMRI}$ was obtained assuming a cuspy density profile inside the SMBH influence radius (see the discussion in Section 2.2) and we impose that $f_{\rm bb}=1$, i.e. that all NCs in the considered mass range host a SMBH.

In addition,  Gair et al. (2004) do not make a fully consistent cosmological integration in a $\Lambda{}$CDM Universe. Furthermore, the number of detectable events reported in table~3 of  Gair et al. (2004) was derived imposing that $z_{\rm max}$ cannot be higher than one.  On the other hand,  Gair et al. (2004) use the same kludge waveforms we adopt (from Barack \&{} Cutler 2004a) and include an accurate treatment for detection statistics (while we consider an averaged signal-to-noise ratio over the duration of the mission). To check the importance of the redshift integration, we can substitute our equation~(\ref{eq:halo}) with a less accurate calculation (see the appendix~C of Mapelli et al. 2010):
\begin{equation}\label{eq:appr}
R=n_{\rm SMBH}(z=0)\,{}\,{}\nu_{\rm EMRI}\,{}\,{}V_c(z_{\rm cut}),
\end{equation}
 where $n_{\rm SMBH}(z=0)$ is the density of SMBHs surrounded by NCs at redshift $z=0$ in the $10^5-6\times{}10^6$ M$_\odot{}$ range, $\nu_{\rm EMRI}$ is defined in Section 2.2, and $V_c(z_{\rm cut})$ is the comoving volume up to redshift $z_{\rm cut}$. For comparison with Gair et al. (2004), we assume that $z_{\rm cut}$ is equal to the minimum between 1 and  $z_{\rm max}$. From equation~(\ref{eq:appr}), we obtain an approximate EMRI detection rate $R\approx{}2\times{}10^3$ yr$^{-1}$ and $R\approx{}2\times{}10^2$ yr$^{-1}$,  for the MH03 and the G12a scaling relation, respectively.
These values are of the same order of magnitude as those listed in our Table~1, indicating that the accurate redshift integration does not affect significantly the rate in the case of LISA.

We now consider the  predictions for NGO, by comparing our results with those from Amaro-Seoane et al. (2012a). In their fig.~21, Amaro-Seoane et al. (2012a) show $z_{\rm max}$ as a function of the SMBH mass, for EMRIs with a 10 M$_\odot{}$ BH. Our Fig.~\ref{fig:fig2} is in fair agreement with the sky-averaged horizons shown in fig.~21 of Amaro-Seoane et al. (2012a), taking into account that  Amaro-Seoane et al. (2012a) sky-averaged horizons were calculated for Teukolsky waveform models (Teukolsky 1973),  $T_{\rm mission}=2$ yr, random generated plunge times $0\le{}t_0/{\rm yr}\le{}5$ and $\langle{\mathrm SNR}\rangle \ge{}20$, whereas in our Fig.~\ref{fig:fig2} we adopt Barack \&{} Cutler (2004a) waveforms, $T_{\rm mission}=t_0=5$ yr and $\langle{\mathrm SNR}\rangle \ge{}10$. The expected EMRI detection rates for NGO, that can be derived from table~2 of Amaro-Seoane et al. (2012a) ($R\approx{}15-25$ yr$^{-1}$) are consistent with the values reported in our Table~2, but quite optimistic for a 2-yr mission.

\section{Conclusions}
NCs seem to be common in galaxies, independently of the Hubble type. Recent studies (Seth et al. 2008a, 2008b; Gonzalez Delgado et al. 2008, 2009; Graham \&{} Spitler 2009) indicate that NCs  co-exist with SMBHs in the same nucleus. The fraction of galaxies hosting both a NC and a SMBH is particularly high among spheroids with stellar mass ranging from $\sim{}10^8{\rm M}_\odot{}$ up to a few $\sim{}10^{10}{\rm M}_\odot{}$.
Given their high stellar densities and high escape velocities, NCs are expected to retain a large fraction of their stellar-mass BHs. 
Thus, it is reasonable to expect that the rate of EMRIs is enhanced in NCs.

 We consider sensitivity curves for three different space-based GW laser interferometric mission concepts: the Laser Interferometer Space Antenna (LISA), the New Gravitational wave Observatory (NGO) and the DECi-hertz Interferometer Gravitational wave Observatory (DECIGO). 

We show that GW signal from SMBH$-$BH mergers can be observed as far as redshift $z\sim{}2$ and $z\sim{}8$, for LISA and DECIGO, respectively, while this limit is substantially lower ($z\sim{}0.3$) for NGO. We predict that, under the most optimistic assumptions, LISA and DECIGO will  detect up to thousands of EMRIs in NCs per year,
 while NGO will  observe up to tens of EMRIs per year.

The estimate of the detection rate  $R$ for  SMBH$-$BH EMRIs is subject to a plethora of severe uncertainties. First, the EMRI rate depends on the dynamics and on the relativistic effects in the neighborhoods ($\lesssim{}10^{-2}$ pc) of the SMBH. In particular, the collisional dynamics of relativistic star clusters around SMBHs is poorly understood. Our predicted rates are upper limits, in the sense that the EMRI rate was derived assuming a mass-segregated cusp in the proximity of the SMBH (Merritt et al. 2011).  The existence of a flat core rather than a mass-segregated cusp in the central parsec of galaxies can severely affect the EMRI rate (e.g., Antonini \&{} Merritt 2012).

Furthermore, the cosmological rate of EMRIs
 depends on the dark-to-luminous mass ratio in galaxies, which is highly uncertain. In addition, we assumed that the fraction of spheroids that host both a SMBH and a NC is $f_{\rm bb}=1$ for a spheroid mass range $10^8\le{}m_{\rm sph}/M_\odot{}\le{}10^{10}$. This is consistent with recent observations (Graham \&{} Spitler 2009), but is still based on a small sample of galaxies ($\sim{}20-30$ galaxies are known to host both a SMBH and a NC in their nucleus). If the available sample suffers from any biases, the actual $f_{\rm bb}$ may be $<1$, although values of $f_{\rm bb}<<1/2$ are unlikely. In addition, $f_{\rm bb}$ might change with redshift. In general, we do not know how NCs and SMBHs evolve with redshift and we can only assume that the halo occupation of NCs and SMBHs is the same up to $z\sim{}2$.

The mass distribution of SMBHs is poorly understood,  especially in low-mass galaxies. Previous studies (e.g.,  Gair et al. 2004; Amaro-Seoane et al. 2012a) derive the SMBH mass distribution from the $m_{\rm SMBH}-\sigma{}$ relation, whereas we use the $m_{\rm SMBH}-m_{\rm sph}$ relation. Both relations are still largely uncertain at the low-mass tail. In the case of the $m_{\rm SMBH}-m_{\rm sph}$ relation, while previous studies proposed a linear scaling (e.g., MH03; H\"aring \&{} Rix 2004), more recent data indicate a steeper scaling ($m_{\rm SMBH}\propto{}m_{\rm sph}^{1.9}$, G12a, G12b). This difference implies large discrepancies (by a factor of 10) in the expected GW detection rate.

Finally, the sensitivity curves of space-borne interferometers will likely be modified in the next few years, as DECIGO is still in the design phase, while the LISA mission concept was recently abandoned and the NGO  mission concept was not selected by the European Space Agency (ESA). 
Differences in the sensitivity curves might strongly affect the predictions that we have computed here.

Therefore, to give an accurate estimate of the detection rate $R$ of EMRIs in NCs is beyond the scope of this paper. On the other hand, our results in Tables~1, 2 and 3 and in Figs.~\ref{fig:fig4} and ~\ref{fig:fig5} can be considered as upper limits, and indicate that EMRIs in NCs might be detected by space-born detectors. In summary, the co-existence of SMBHs and NCs can boost significantly our chance of observing GWs from EMRIs in the galactic centres, although more studies of the dynamics of relativistic star clusters are needed, to quantify this with accuracy.

\section*{Acknowledgments}
We thank the anonymous referee for the accurate and incisive comments, which helped us improving the manuscript. We thank D.~Merritt for his helpful comments, M.~Colpi and A.~Klein for stimulating discussions. 
We thank Wayne~Hu for making his transfer function code publicly available.
A.W.Graham was supported under the Australian Research Council's Discovery
Projects funding scheme (DP110103509).

\onecolumn
\appendix
\section{Method to estimate $z_{\rm max}$}
 We define $z_{\rm max}(m_{\rm SMBH},\,{}m_{\rm co})$ as the maximum redshift at which an event can be detected  with a sky location and orientation averaged
 signal-to-noise ratio $\langle {\rm SNR}\rangle\ge{}10$ by a 
 single interferometer. Mergers of SMBHs with stellar BHs are classified as extreme mass-ratio inspirals (EMRIs). For such systems, 
  one can estimate $\langle {\rm SNR}\rangle$ following the approach and waveform approximation described by Barack \&{} Cutler (2004a). The angle-averaged signal-to-noise ratio $\langle {\rm SNR}\rangle$ for a single synthetic two-arm Michelson detector is given by:
\begin{equation}
\langle {\rm SNR^2}\rangle=\sum_n\int{}\frac{h_{c,n}^2(f_n(t))}{5\,{}f_n(t)\,{}S_h(f_n(t))}\,{}{\rm d}(\ln{f_n(t)}),
\end{equation}
where $f_n(t)$ indicates  
 the frequency of the different harmonics at multiple integers of the orbital frequency
(in our calculations, we adopt $n=1,...,20$, which 
 safely accounts for all the GW power)
and is defined to be $f_n(t)\equiv{}n\,{}\nu{}(t)+\dot{\tilde{\gamma{}}}(t)/\pi{}$ (where $\nu{}(t)$ is the orbital frequency and $\tilde{\gamma{}}(t)$ is the direction of pericentre with respect to $\hat{L}\times{}\hat{S}$ -i.e. the vector product 
 of the orbital angular momentum unit vector $\hat{L}$ and the SMBH spin direction $\hat{S}$).  
$h_{c,n}(f_n(t))$ is the characteristic amplitude and has been derived following equation (56) of Barack \&{} Cutler (2004a; see also Cutler et al. 1993, 1994). Finally, $S_h(f_n(t))$ accounts for the total noise of the interferometer and 
 we give expressions for LISA, NGO and DECIGO in the next subsection.
In this Paper, we derive $f_n(t)$ by solving, backwards in time, the system of differential post-Newtonian equations (27-30) of Barack \&{} Cutler (2004a). In particular, we evolve the mean anomaly $\Phi{}(t)$, the orbital frequency $\nu{}(t)$, the direction of pericentre $\tilde{\gamma{}}(t)$ and the orbital eccentricity $e(t)$ starting from the LSO, backwards in time down to the assumed starting time of the observation. We define as $t_0$ the duration elapsed between the starting time of the observation and the epoch when the companion reaches the LSO (in this paper, we consider four cases: $t_0=5,\,{}6,\,{}10$ and 15 yr). We furthermore define $T_{\rm mission}$ the duration of the mission (in this paper, we consider two cases:  $T_{\rm mission}=5$ and 2 yr).
 As arbitrary initial conditions, we set $\Phi{}_{\rm LSO}=0$,  $\tilde{\gamma{}}_{\rm LSO}=0$, and we keep the LSO eccentricity $e_{\rm LSO}$ as a free parameter (see Table~1), which yields $\nu{}_{\rm LSO}=\left(2\,{}\pi{}m_{\rm SMBH}\right)^{-1}\,{}\left[\left(1-e^2_{\rm LSO}\right)/\left(6+2\,{}e_{\rm LSO}\right)\right]^{3/2}$.

Using the above equations, we can derive the luminosity distance $D_L(z_{\rm max}(m_{\rm SMBH},\,{}m_{\rm co}))$ at which an event can be detected  with a ${\rm SNR}\ge{}10$ by a certain interferometer (see, e.g., Mapelli et al. 2010 and references therein). 
We can, thus, derive $z_{\rm max}(m_{\rm SMBH},\,{}m_{\rm co})$ by inverting the expression of the luminosity distance in  a flat $\Lambda{}$CDM model:
\begin{eqnarray}
D_L(z_{\rm max}(m_{\rm SMBH},\,{}m_{\rm co}))=\frac{c}{H_0}(1+z_{\rm max}(m_{\rm SMBH},\,{}m_{\rm co}))\,{}
\int_0^{z_{\rm max}{(m_{\rm SMBH},\,{}m_{\rm co})}}{\frac{{\rm d}z}{{\mathcal E}(z)}}.
\label{e:DL}
\end{eqnarray}

\subsection{LISA noise}\label{sec:LISA}
$S_h(f_n(t))$ accounts for the total noise of LISA noise and is defined as (equation 54 of Barack \&{} Cutler 2004a):
\begin{equation}
S_h(f_n(t))=S_h^{\rm inst+gal}(f_n(t))+S_h^{\rm ex gal}(f_n(t)),
\end{equation}
where $S_h^{\rm inst+gal}(f_n(t))$ is defined in equation (52) of Barack \&{} Cutler (2004a, see also Hughes 2002) and accounts for both the instrumental noise ($S_h^{\rm inst}(f_n(t))$, Finn \&{} Thorne 2000) and the Galactic confusion noise ($S_h^{\rm gal}(f_n(t))$, Nelemans, Yungelson \&{} Portegies Zwart 2001, see also Bender \&{} Hils 1997):
\begin{equation}
S_h^{\rm inst+gal}(f_n(t))=\textrm{min}\left[\frac{S_h^{\rm inst}(f_n(t))}{\exp{(-\kappa{}\,{}T_{\rm mission}^{-1}\,{}N_f)}},\,{}S_h^{\rm inst}(f_n(t))+S_h^{\rm gal}(f_n(t))\right],
\end{equation}
where $\kappa{}\sim{}4.5$ (Cornish 2003), $T_{\rm mission}$ yr is the duration of the mission,  $N_f=2\times{}10^{-3}\textrm{Hz}^{-1}(\textrm{Hz}/f_n(t))^{11/3}$ (Hughes 2002). $S_h^{\rm inst}(f_n(t))$ is the same as in equation (48) of Barack \&{} Cutler (2004a). $S_h^{\rm gal}(f_n(t))$ is the same as in equation (51) of  Barack \&{} Cutler (2004a).
Finally, $S_h^{\rm ex gal}(f_n(t))$ accounts for the extragalactic white dwarf background (Farmer \&{} Phinney 2003) and is the same as in  equation (50) of Barack \&{} Cutler (2004a).
\subsection{NGO noise}\label{sec:NGO}
 In the case of NGO, $S_h^{\rm inst}(f_n(t))$ is defined as  (equation 5 of Amaro-Seoane et al. 2012a):
\begin{equation}
S_h^{\rm inst}(f_n(t))=\frac{20}{3}\,{}\frac{4\,{}S_{\rm x,acc}(f_n(t))+S_{\rm x,sn}(f_n(t))+S_{\rm X,omn}(f_n(t))}{L^2}\,{}\left\{1+\left[\frac{f_n(t)}{0.41\,{}\left(\frac{c}{2\,{}L}\right)}\right]^2\right\},
\end{equation}
where $L=10^9$ m, $S_{\rm x,acc}(f_n(t))=1.37\times{}10^{-32}\left(1+\frac{10^{-4}{\rm Hz}}{f_n(t)}\right)\,{}\left(\frac{\rm Hz}{f_n(t)}\right)^4$ m$^{2}$ Hz$^{-1}$ is the power spectral density of the residual acceleration of the test mass, $S_{\rm x,sn}(f_n(t))=5.25\times{}10^{-23}$ m$^2$  Hz$^{-1}$ is the shot noise power spectral density, and $S_{\rm X,omn}(f_n(t))=6.28\times{}10^{-23}$ m$^2$  Hz$^{-1}$ is the   power spectral density of the other possible measurement noises.

As in the case of LISA, we combine the instrumental noise $S_h^{\rm inst}(f_n(t))$ with the Galactic confusion noise and with the extragalactic white dwarf background (see Section A.1). On the other hand, the  Galactic confusion noise and the extragalactic white dwarf background are negligible compared to $S_h^{\rm inst}(f_n(t))$ in the case of NGO.
\subsection{DECIGO noise}
In the case of DECIGO, we calculate the noise as in equation~(1) of Yagi \&{} Tanaka (2010):
\begin{equation}
S_h(f_n(t))=\textrm{min}\left[\frac{S_h^{\rm inst}(f_n(t))}{\exp{(-\kappa{}\,{}T_{\rm mission}^{-1}\,{}N_f)}},\,{}S_h^{\rm inst}(f_n(t))+S_h^{\rm gal}(f_n(t)){\mathcal R}(f_n(t))\right]+S_h^{\rm ex gal}(f_n(t)){\mathcal R}(f_n(t))+0.01\,{}S_h^{\rm NS}(f_n(t)),
\end{equation}
where the instrumental noise spectral density ($S_h^{\rm inst}(f_n(t))$) is given by (Yagi \&{} Tanaka 2010):
\begin{equation}
S_h^{\rm inst}(f_n(t))=5.3\times{}10^{-48}\left[(1+x^2)+\frac{2.3\times{}10^{-7}}{x^4(1+x^2)}+\frac{2.6\times{}10^{-8}}{x^4}\right]\,{}\textrm{Hz}^{-1},
\end{equation}
where $x=f_n(t)/f_p$ with $f_p\equiv{}7.36$ Hz. $\kappa{}$, $T_{\rm mission}$, $N_f$, $S_h^{\rm gal}(f_n(t))$ and $S_h^{\rm ex gal}(f_n(t))$ are the same as defined for LISA in Section~\ref{sec:LISA}. Finally, the factor ${\mathcal R}(f_n(t))\equiv{}\exp{\left[-2\,{}(f_n(t)/0.05\textrm{ Hz})^2\right]}$ represents the cutoff of the white dwarf$-$white dwarf binary confusion noise (Yagi \&{} Tanaka 2010).

\section{Estimate of $p$ from the Press-Schechter formalism}
We can derive $p$ in the following way.  Driver et al. (2007) estimate that the current stellar mass density of spheroids is $\rho{}_{\rm sph}\sim{}3.0\times{}10^8h\,{}{\rm M}_\odot{}\,{}{\rm Mpc}^{-3}$ (where $h=0.71$, Larson et al. 2011). Using the Schechter function (see Driver et al. 2007 for details),
 we find that the mass fraction of spheroids with mass $10^8\le{}m_{\rm sph}/{\rm M}_\odot{}\le{}10^{10}$ is $g\sim{}0.34$. Thus, the current mass density of halos hosting spheroids with mass $10^8\le{}m_{\rm sph}/{\rm M}_\odot{}\le{}10^{10}$ is

\begin{equation}\label{eq:norm}
\rho{}_{\rm h}= \frac{g\,{}\rho{}_{\rm sph}}{f_\ast{}}\,{}\left(\frac{\Omega{}_{\rm M}}{\Omega{}_{\rm b}}\right)\,{}\left(\frac{\rho{}_{\rm sph}}{\rho{}_{\rm sph}+\rho_{\rm disc}}\right)^{-1},
\end{equation}
where $f_\ast{}$ is the fraction of baryons in stars, $\rho_{\rm disc}$ is the current stellar mass density of discs, $\Omega{}_{\rm M}=0.27$ and $\Omega{}_{\rm b}=0.044$ (Larson et al. 2011). 
Assuming $g=0.34$, $f_\ast{}=0.119\,{}h$ and $\rho_{\rm disc}=4.4\times{}10^8h\,{}{\rm M}_\odot{}\,{}{\rm Mpc}^{-3}$ (from Driver et al. 2007), we find $\rho{}_{\rm h}=1.3\times{}10^{10}\,{}{\rm M}_\odot{}\,{}{\rm Mpc}^{-3}$.

A density $\rho{}_{\rm h}=1.3\times{}10^{10}\,{}{\rm M}_\odot{}\,{}{\rm Mpc}^{-3}$ can be obtained by integrating the Press-Schechter function (Press \&{} Schechter 1974; Eisenstein \&{} Hu 1998, 1999) between $m_{\rm h1}=9.44\times{}10^9\,{}{\rm M}_\odot{}$ and $m_{\rm h2}=9.44\times{}10^{11}\,{}{\rm M}_\odot{}$ (which satisfies the further requirement that $m_{\rm h2}=100\,{}m_{\rm h1}$). $m_{\rm h1}=9.44\times{}10^9\,{}{\rm M}_\odot{}$ and $m_{\rm h2}=9.44\times{}10^{11}\,{}{\rm M}_\odot{}$  can be used as integration limits in equation~(\ref{eq:halo}). Finally, we can derive also the value of $p$ as $p=m_{\rm sph1}/m_{\rm h1}=1.06\times{}10^{-2}$.

 We note that this value of $p$ matches very well with the observations of the Milky Way, but there may be large deviations from it, depending on the morphological type (see, e.g., Graham \&{} Worley 2008). A different value of $p$ may significantly affect the results (see the  examples in Tables~1, 2 and 3).

\end{document}